\documentclass[JHEP,12pt]{article}

\usepackage{graphics,color,soul}
\usepackage{graphicx}
\usepackage{amssymb}
\usepackage[export]{adjustbox}
\usepackage{jheppub}
\usepackage{lmodern,mathtools}
\usepackage{caption}
\usepackage{subcaption}
\usepackage{booktabs}
\usepackage[english]{babel}
\usepackage{amsmath,amssymb,amsbsy,amstext, amsthm, simplewick}
\usepackage{hyperref}
\usepackage{tikz}
\usetikzlibrary{decorations.markings,decorations.pathmorphing, patterns,shapes}
\usepackage{xcolor}
\usepackage[normalem]{ulem}

\usepackage{caption}

\usepackage{amsfonts}
\usepackage{amssymb}
\usepackage{upgreek}
\usepackage{simplewick}
 \usepackage{exscale,relsize}
\usepackage{mathtools}
\usepackage{comment}

\usepackage[margin=1cm,labelfont={sf,bf,scriptsize},textfont={sf,scriptsize}]{caption}

\def\be#1\ee{\begin{align}#1\end{align}}

\tikzset{
  cross/.style={
    line width=0.8pt,
    cross out,
    draw=black,
    minimum size=4pt,
  }
}

\begin{document}

\title{Bouncing off a stringy singularity
}

\author[1]{Matthew Dodelson,}
\author[2]{Cristoforo Iossa,}
\author[3]{and Robin Karlsson}

\affiliation[1]{Center for the Fundamental Laws of Nature, Harvard University, Cambridge, MA, USA}
\affiliation[2]{Section de Mathématiques, Université de Genève, 1211 Genève 4, Switzerland}
\affiliation[3]{Mathematical Institute, University of Oxford, Andrew Wiles Building, Radcliffe Observatory Quarter, Woodstock Road, Oxford, OX2 6GG, UK}

\abstract{A sharp signature of the black hole singularity in holography is a divergence in the boundary thermal two-point function at a specific point in the complex time plane. This divergence arises from a null geodesic that bounces off the black hole singularity. At finite 't Hooft coupling, stringy corrections to the bulk dynamics cannot be neglected, and the fate of the bouncing geodesic is an open question. We propose a simple scenario in which the singularity in the two-point function is shifted slightly into the complex plane, thereby smoothing it out into a finite-size bump. We demonstrate this smoothing explicitly in a microscopic example, namely the Sachdev-Ye-Kitaev model at infinite temperature, where the correlator is under analytic control. Our result suggests a bulk description of planar theories at finite coupling as stringy black \nolinebreak holes.
}

\begin{flushleft}
\end{flushleft}

\maketitle
\section{Introduction}
\begin{figure}[t]
\begin{subfigure}{0.45\textwidth}
\centering
\begin{tikzpicture}[x=1.2cm, y=1cm]
    \def\boundaryX{2}   
    \def\boundaryY{2.5}  
    \def\singularityBend{0.75} 
    \draw[thick] (-\boundaryX, -\boundaryY) -- (-\boundaryX, \boundaryY); 
    \draw[thick] (\boundaryX, -\boundaryY) -- (\boundaryX, \boundaryY); 
    \draw[thick,decorate, decoration={snake, amplitude=1.5pt, segment length=6pt}]
        (-\boundaryX, \boundaryY)--(-\boundaryX+0.15, \boundaryY)  .. controls +(0, -\singularityBend) and +(0, -\singularityBend) .. (\boundaryX-0.15, \boundaryY)--(\boundaryX, \boundaryY);
    \draw[thick, decorate, decoration={snake, amplitude=1.5pt, segment length=6pt}]
        (-\boundaryX, -\boundaryY)--(-\boundaryX+0.15, -\boundaryY) .. controls +(0, \singularityBend) and +(0, \singularityBend) .. (\boundaryX-0.15, -\boundaryY)--(\boundaryX, -\boundaryY);
    \draw[gray] (-\boundaryX, -\boundaryY) -- (\boundaryX, \boundaryY);
    \draw[gray] (-\boundaryX, \boundaryY) -- (\boundaryX, -\boundaryY);
    \node at (2.5,0.1) {\small $t=0$};
    \node[circle, fill, inner sep=1.3pt] at (2,0) {};
    \draw[red,thick] (-2,-0.6)--(0,2);
    \draw[red,thick] (0,2)--(2,-0.6);
    \draw[gray, dashed] (-2,0) -- (2,0);
\end{tikzpicture}
\caption{\label{fig:bouncingPhoto}}
\end{subfigure}
\begin{subfigure}{0.45\textwidth}
\centering
\begin{tikzpicture}[x=1.2cm, y=1cm]
    \def\boundaryX{2}   
    \def\boundaryY{2.5}  
    \def\singularityBend{0.75} 
     \def\xmin{-2}
  \def\xmax{2}
    \draw[thick] (-\boundaryX, -\boundaryY) -- (-\boundaryX, \boundaryY); 
    \draw[thick] (\boundaryX, -\boundaryY) -- (\boundaryX, \boundaryY); 
    \draw[thick, decorate, decoration={snake, amplitude=1.5pt, segment length=6pt}]
        (-\boundaryX, \boundaryY)--(-\boundaryX+0.15, \boundaryY)  .. controls +(0, -\singularityBend) and +(0, -\singularityBend) .. (\boundaryX-0.15, \boundaryY)--(\boundaryX, \boundaryY);
    \draw[thick, decorate, decoration={snake, amplitude=1.5pt, segment length=6pt}]
        (-\boundaryX, -\boundaryY)--(-\boundaryX+0.15, -\boundaryY) .. controls +(0, \singularityBend) and +(0, \singularityBend) .. (\boundaryX-0.15, -\boundaryY)--(\boundaryX, -\boundaryY);
    \draw[gray] (-\boundaryX, -\boundaryY) -- (\boundaryX, \boundaryY);
    \draw[gray] (-\boundaryX, \boundaryY) -- (\boundaryX, -\boundaryY);
      \fill[red!20]
    plot[domain=\xmin:\xmax, samples=100] (\x, {-.5*\x*\x +1.4})
    -- plot[domain=\xmax:\xmin, samples=100] (\x, {-.625*\x*\x +1.9})
    -- cycle;
    \draw[gray, dashed] (-2,0) -- (2,0);
\end{tikzpicture}
\caption{\label{worldsheet}}
\end{subfigure}
\centering
\caption{(a) A bouncing geodesic in the maximally extended $\text{AdS}_5$-Schwarzschild black brane. In order for the geodesic to meet the singularity at the center of the diagram it must start with boundary time $t=-\frac{\beta}{4}<0$. (b) A bouncing worldsheet in string theory.
}
\end{figure}
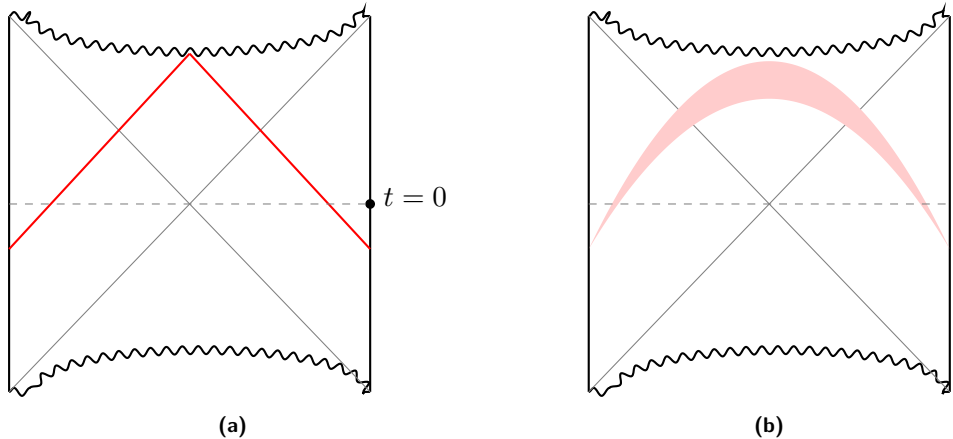

\indent In a theory of point particles, curvature singularities signal a breakdown of physics. Yet in string theory, such singularities can be softened or even resolved by the extended nature of strings. One classic example arises in the study of orbifolds \cite{Dixon:1985jw}, which appear singular to a point particle. When strings propagate on these spaces, the extended structure of the string allows it to “feel” twisted sectors and new degrees of freedom that render scattering processes finite and well-behaved. This phenomenon motivates the study of stringy geometry, a framework in which the geometric notions of smoothness and singularity are reinterpreted through the lens of string dynamics.\\
\indent The advent of AdS/CFT \cite{Maldacena:1997re,Gubser:1998bc,Witten:1998qj} introduced a new perspective on stringy geometry. At infinite ’t Hooft coupling, boundary correlators develop nontrivial singularities whenever points are connected by lightlike trajectories through the bulk, providing a sharp probe of the bulk spacetime. Examples include the bulk point singularity \cite{Gary:2009ae,Maldacena:2015iua} in vacuum correlators and bulk cone singularities \cite{Hubeny:2006yu,Dodelson:2020lal,Dodelson:2023nnr} in nontrivial excited states. As one moves away from the infinite coupling limit, stringy corrections smooth out these singularities \cite{Maldacena:2015iua,Dodelson:2020lal}, signaling that local geometry emerges only in the supergravity regime. See \cite{Gesteau:2023rrx,Ouseph:2023juq,Gesteau:2024rpt,Belin:2025nqd,Herderschee:2025nsb} for other perspectives on stringy geometry in AdS/CFT. \\
\indent The most dramatic type of curvature singularity lies at the center of a black hole, and it is natural to ask how this singularity appears to a stringy probe \cite{Horowitz:1989bv,Martinec:1994xj,Zigdon:2024ljl,Shahbazi-Moghaddam:2025fxi}. The two-sided Schwarzschild black hole was studied in \cite{Kraus:2002iv,Fidkowski:2003nf,Festuccia:2005pi}, where it was shown that suitably analytically continued boundary correlators develop singularities associated with geodesics that bounce off the (spacelike) black hole singularity, as illustrated in Figure \ref{fig:bouncingPhoto}. More recently, these bouncing geodesics were shown to govern the large real-frequency behaviour of the retarded thermal correlator \cite{Afkhami-Jeddi:2025wra} through a bulk WKB analysis. From the CFT perspective, these singularities were identified in \cite{Ceplak:2024bja} as arising from the multi-trace stress tensor operators $[T_{\mu\nu}^k]$ as $k\to\infty$. This geometric description in terms of bulk geodesics is expected to hold in the semiclassical regime of small Newton's constant $G_N$ (corresponding to a large number of degrees of freedom $N^\#\sim 1/G_N$) and small string length $\sqrt{\alpha'}$ in units of the AdS radius (or large 't Hooft coupling $\lambda$).

 In this paper, we explore the fate of the bouncing geodesic singularity in large $N$ theories away from the limit of infinite coupling. Heuristically, the bouncing geodesic expands into a worldsheet as in Figure \ref{worldsheet}. We first consider the point particle limit of AdS/CFT and derive the analytic structure of thermal two-sided correlators in the complex time domain. The result is shown to follow from the asymptotic spectrum of quasinormal modes (QNMs) using the thermal product formula \cite{Dodelson:2023vrw}. A crucial ingredient in that work was the observation that there are no zeroes in the complex frequency plane. Fourier transforming to time, this results in a lattice of singularities, with the most important feature being that all singularities lie on what we call \emph{real sections}. For AdS-Schwarzschild, the real sections are given by $\text{Im}\, t =\pm m\beta/2$ with $m=1,2,\ldots$. For $m=1$ this is the OPE singularity, while for $m=2,3,\ldots$ these are due to bouncing geodesics. 

Since the bulk geometry becomes strongly curved near the black hole singularity, we are not able to analyze the worldsheet in Figure \ref{worldsheet} explicitly, in contrast with the situation for the bulk point and bulk cone singularities. Instead we take the strategy of looking for a boundary model that exhibits the desired effects. It turns out that an ideal candidate is the SYK model \cite{Maldacena:2016hyu,kitaev,Sachdev_1993,Polchinski:2016xgd} at infinite temperature. There is a sense in which this model has a small string scale, so that we can directly observe what happens to the bouncing geodesic singularities at large but finite coupling. By computing the correlator to high accuracy and analytically continuing, we will find that the singularities in the correlator move slightly off the real sections into the complex time plane. In other words, the bouncing geodesic singularities turn into bumps of finite height, in accordance with the general philosophy of stringy geometry.  \\
\indent This work is organized as follows. In Section \ref{sec:Holography}, we explain how the bouncing singularities in the complex time plane arise succinctly from QNMs and the thermal product formula \cite{Dodelson:2023vrw}. We also analyze how the structure of singularities changes when charge is added to the black hole, so that the singularity becomes timelike. In Section \ref{sec:zeroes}, we explore the consequences of zeroes in complex frequency for the bouncing singularities, finding that the addition of zeroes spoils the lattice structure of the singularities.  This is motivated in part by what happens in SYK away from the maximally chaotic regime, which we explore in Section \ref{sec:SYK}. We conclude in Section \ref{outlook} with some open problems.\\
\indent The paper \cite{ceplak} contains some overlap with Section \ref{chargedsec} of this work, and will appear simultaneously on the arXiv.
\section{Bouncing geodesics from quasinormal modes}\label{sec:Holography}
In this section we explain how geodesics bouncing off black hole singularities are imprinted onto boundary correlators in holography. The basic strategy is to compute the correlator at high frequencies using the thermal product formula \cite{Dodelson:2023vrw} and then Fourier transform. As we will see, the singularities in complex time form a lattice whose basis vectors are determined by the asymptotics of the QNMs.

We start by reviewing some basic facts about singularities induced in holographic correlators by null geodesics bouncing off the black hole singularity. Let us consider the maximally extended Schwarzschild black brane geometry in $\text{AdS}_{5}$, given by the metric 
\be\label{eq:metric}
ds^2 = -f(r)\, dt^2+\frac{dr^2}{f(r)}+r^2\, d\Omega^{2}_{3}, \hspace{10 mm} f(r) = r^2-\frac{\mu}{r^{2}}.
\ee 
In \cite{Fidkowski:2003nf} it was shown that the Penrose diagram of \eqref{eq:metric} is not a square, since null radial geodesics starting from the left and right boundaries at boundary time $t=0$ do not meet at the singularity. 
In order for such geodesics to meet at the singularity they must start with a nonzero boundary time separation. These are commonly called \emph{bouncing geodesics} since they can be thought of as bouncing off the singularity, as depicted in Figure \ref{fig:bouncingPhoto}. 

Bouncing geodesics connect two points on the boundaries by null lines and therefore are expected to produce singularities in suitable analytic continuations of the correlator at certain points in the complex time domain. We refer the reader to Appendix \ref{app:ancont} for a discussion of the general analytic structure of thermal correlators. The location of the singularities can be computed as follows. The redshift factor $f(r)$ generically has a simple zero at the horizon $r=r_s$,
\be\label{eq:redshift}
f(r) = \frac{4\pi}{\beta}(r-r_S)+{O}((r-r_S)^2).
\ee 
The time along a radial null geodesic starting at $t=0$ at the AdS boundary and ending at a point $r$ in the bulk is given by 
\be \label{tofr}
t(r) = \int_{r}^\infty\frac{dr'}{f(r')}. 
\ee
Note that the integrand contains a simple pole at $r=r_S$ due to \eqref{eq:redshift}. Under a suitable $i\epsilon$ prescription, this leads to a relative shift $\beta/4$ between $\text{Im }t(0)$ and $\text{Im }t(\infty)$. A bouncing geodesic as in Figure \ref{fig:bouncingPhoto} crosses the horizon twice, connecting points separated by boundary time
\be
\Delta t = \frac{(1+i)\beta}{2} \,.
\ee
\indent More generally, we can have geodesics which bounce $n$ times off the singularity in the maximally extended black hole, and thus go $2n$ times around the poles at the horizon \cite{Amado:2008hw}. Accordingly, the time separation is given by
\be\label{eq:GeometricSection}
\Delta t_n =  \frac{(\pm 1+i)n\beta}{2},
\ee 
where the $\pm$ depends on if the geodesic bounces clockwise or counterclockwise. For a general black hole, let us define the \emph{real sections} as the set of all complex $t$ whose imaginary part matches the imaginary part of (\ref{tofr}) with an appropriate choice of $r$ contour which is real away from the horizons. The set of real sections for the black brane is then given by $\text{Im} \, t=n\beta/2$ for integer $n$. Since the time separation of null geodesics can acquire an imaginary part only by crossing horizons, all singularities of boundary correlators dual to an AdS black hole with a spacelike singularity should lie on these real sections. In the next subsection, we will show how this claim follows from the asymptotics of QNMs along with the thermal product formula \cite{Dodelson:2023vrw}, and discuss several basic examples. While the arguments above rely on the geodesic approximation, we will see below that the bouncing singularities are present even at finite $\Delta$ (see also \cite{Ceplak:2024bja,Afkhami-Jeddi:2025wra}). 

\subsection{Singularities in complex time from QNMs}
Let us now show how bouncing singularities arise from QNMs in holographic correlators. Rather than restricting to specific examples, we will show that the result takes a universal form for all spherically symmetric nonextremal AdS black holes, depending only on the asymptotic structure of QNMs. We consider the two-sided correlator $C(\omega)$, whose Fourier transform $C(t)$ is defined in (\ref{cdef}). Here we are suppressing dependence on the spatial momentum. By working at fixed spatial momentum we are integrating over space, which affects the analytic structure in complex time as discussed in Appendix \ref{app:ancont}. In \cite{Dodelson:2023vrw} it was shown that $C(\omega)$ can be written as a product over its poles,
\begin{equation}\label{eq:productformula}
    C(\omega) = \frac{C(0)}{\prod_{n=1}^\infty \left(1-\frac{\omega^2}{\omega_n^2}\right)\left(1-\frac{\omega^2}{(\omega_n^*)^2}\right)}.
\end{equation}
Given a QNM at $\omega_n$, it follows from the properties of $C$ that there are also poles at $-\omega_n,\omega_n^*,$ and $-\omega_n^*$. Imaginary simple poles contribute with a factor $1-\omega^2/\omega_n^2$ in the denominator.

Asymptotically in the mode number $n\gg1$ one generically finds multiple families of QNMs parameterized by an integer $n$,
\be\label{eq:qnmParameter}
\omega_n = re^{i\theta} n +se^{i\phi}+\ldots,
\ee 
where ellipses denote subleading corrections as $n\to\infty$. Let us focus on the case of a single line of QNMs. In this case the consistency conditions derived in \cite{Dodelson:2023vrw} lead to the following relations,
\begin{align}
\beta &= \frac{4\pi \sin\theta}{r},\label{eq:betaDelta}\\
2\Delta-d &= \frac{4s\cos(\theta-\phi)}{r}+2.
\end{align}
Inserting the modes \eqref{eq:qnmParameter} into the product formula \eqref{eq:productformula} and ignoring subleading corrections leads to
\be\label{eq:g12gd}
C(\omega) \approx  \prod_\pm \Gamma\left(\frac{e^{i\theta}}{r}\left( \pm \omega+e^{-i\phi}s\right)+1\right) \Gamma\left(\frac{e^{-i\theta}}{r}\left( \pm \omega+e^{i\phi}s\right)+1\right),\hspace{5 mm}|\omega|\to \infty.
\ee
There is an arbitrary normalization factor in this equation, which we ignore from now on.\\
\begin{figure}[t]
\centering
\begin{tikzpicture}[scale=1.1]
\fill[color=gray!20] (-4.5,-1) rectangle (4.5,1);

  \draw[->] (-4.5,0) -- (4.5,0) node[right] {$\text{Re}\,t$};
  \draw[->] (0,-1) -- (0,4.5) node[above] {$\text{Im}\,t$};
 
    \draw[dashed,gray, thick] (-4.5,1) -- (4.5,1);
   \draw[dashed,gray, thick] (-4.5,-1) -- (4.5,-1);
       \draw[dashed,gray, thick] (-4.5,2) -- (4.5,2);
    \draw[dashed,gray, thick] (-4.5,3) -- (4.5,3);
        \draw[dashed,gray, thick] (-4.5,4) -- (4.5,4);
 
  \coordinate (z0) at (0,1);
  \filldraw[black] (z0) circle (2pt) node[below right] {$\frac{i\beta}{2}$};
  \filldraw[black] (0,-1) circle (2pt) node[above right] {$-\frac{i\beta}{2}$};

  \foreach \m in {0,1,2,3,5,6,7} {
    \foreach \n in {0,1,2,3,5,6,7} {
      \pgfmathsetmacro{\sum}{\m+\n}
      \ifdim \sum pt < 4pt
        \pgfmathsetmacro{\x}{1*(\n - \m)}
        \pgfmathsetmacro{\y}{1 + 1*(\n + \m)}
        \filldraw[black] (\x,\y) circle (2pt);
      \fi
    }
  }
  \draw[->, ultra thick, red] (z0) -- ++(1,1) node[right, black] {};
  \draw[->, ultra thick, red] (z0) -- ++(-1,1) node[above left, black] {};  
  \node at (1.1,1.5) {$v_+$};
  \node at (-1.1,1.5) {$v_-$};

\end{tikzpicture}
\caption{Lattice of singular points of $C(t)$ in complex time, with real sections shown as dashed lines. The black dots denote singularities at the lattice points (\ref{latticedef}), while the red arrows are the vectors generating the lattice. In the lower half plane there are singularities at $t=-t_{nm}$.}
\label{fig:lattice1}
\end{figure}
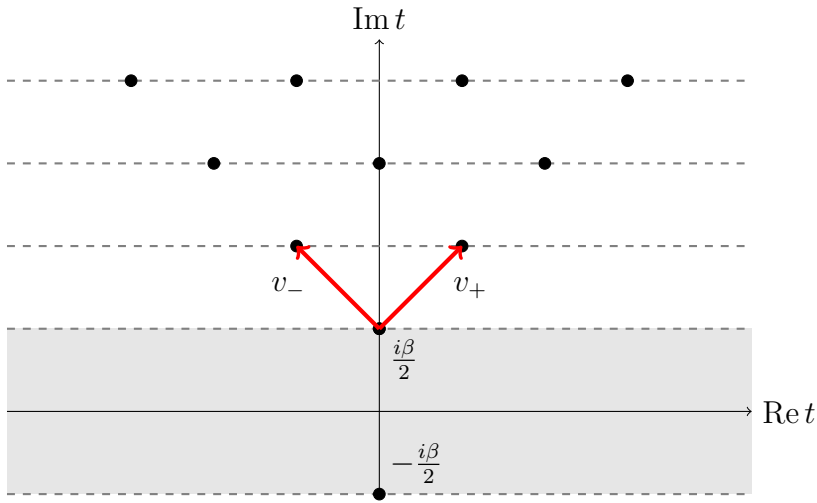
\indent This formula captures the asymptotics of $C(\omega)$ as $|\omega|\to \infty$ along any direction in the complex $\omega$ plane. As we will briefly discuss later, subleading orders do not affect the leading behavior close to the singularities in the time domain. To obtain the behavior at large real $\omega$, we use the reflection formula to transform the expression into a form where the arguments of all gamma functions are positive and then apply Stirling's formula. This gives
\begin{align}\label{capprox}
C(\omega)\approx \frac{\omega^{2\Delta-d}}{\sin\left(\frac{\pi e^{-i\theta}}{r}(\omega-e^{i\phi}s)\right)\sin\left(\frac{\pi e^{i\theta}}{r}(\omega-e^{-i\phi}s)\right)}.
\end{align}
This equation should be viewed as a transseries \cite{https://doi.org/10.1002/zamm.19740540718}, where we keep the leading perturbative term around each exponentially small contribution from the sines in the denominator. We will comment on potential subtleties with this prescription later.\\
\indent We now perform the Fourier transform to the time domain. We will start with real time, and then analytically continue the answer beyond the strip of analyticity. The Fourier transform is
\be 
C(t) &=\int_{0}^{\infty} \frac{d\omega}{\pi}\,  \cos(\omega t)C(\omega),
\ee 
which can be computed by expanding the sines in the denominator of (\ref{capprox}) into a geometric series and integrating term by term. We find
\be\label{eq:resn}
C(t) &\approx \sum_{n,m=0}^{\infty}\frac{e^{\frac{2\pi is}{r}(me^{i(\phi-\theta)}-ne^{i(\theta-\phi)})}}{(i(t-t_{nm}))^{2\Delta-d+1}}+(t\to -t).
\ee
As depicted in Figure \ref{fig:lattice1}, this equation has singularities on the lattice
\begin{align}\label{latticedef}
t_{nm} = \frac{i\beta}{2}+ n v_+ +m v_-, \,\hspace{10 mm} n,m \in \mathbb{Z}_{\ge0},
\end{align}
where the lattice basis vectors are given by $v_{\pm}=\pm 2\pi e^{\pm i\theta}/r$. The reflected lattice points $-t_{nm}$ are also singular. Note that
\be\label{eq:SingsOnReals}
\text{Im }t_{nm} = \frac{\beta}{2}(n+m+1),
\ee
so that all singularities lie on real sections.
For generic $\Delta$, these are branch cut singularities. 

For concreteness, let us now discuss several examples. First consider a scalar of dimension $\Delta$ in an $\text{AdS}_{d+1}$ black brane, for which the QNMs are asymptotically given by \cite{Natario:2004jd,Cardoso:2004up}
\begin{align}\label{braneas}
\omega_n \approx \frac{4\pi}{\beta}\sin\left(\frac{\pi}{d}\right)e^{\frac{i\pi}{d}}\left(n+\frac{\Delta}{2}-\frac{d+2}{4}-\frac{i\log 2}{2\pi}\right).
\end{align}
For these values of the asymptotic parameters, the relative normalization between the singularity at $t_{nm}$ and the OPE singularity becomes 
\begin{align}
e^{\frac{2\pi is}{r}(me^{i(\phi-\theta)}-ne^{i(\theta-\phi)})}=2^{m+n}i^{(2\Delta-d-2)(m-n)}.
\end{align}
This matches the result of \cite{Afkhami-Jeddi:2025wra} in the case $n=1,m=0$.

For certain values of the parameters, the gamma functions in (\ref{eq:g12gd}) simplify and we can perform the Fourier transform exactly. For example, consider the known case of R-currents at zero spatial momentum in $\mathcal{N}=4$ SYM in $d=4$ \cite{Myers:2007we}. One has $\Delta = 3$, and we set $\beta=2\sqrt{2}\pi$ for simplicity. This gives $r=1, \theta=\pi/4$ and $s=0$, and the spectral function (\ref{eq:g12gd}) simplifies to
\be\label{cspecial}
C(\omega)=\frac{\omega^2}{\sin(\pi e^{\frac{i\pi}{4}}\omega)\sin(\pi e^{-\frac{i\pi}{4}}\omega)}.
\ee
The Fourier transform can be performed by closing the contour in the lower half plane and picking up the poles. This gives
\begin{equation}
\label{eq:geng12k}
    C(t) =  \sum_{k=0}^{\infty} \frac{e^{\frac{i\pi}{4}}\sinh\left(\frac{e^{\frac{i\pi}{4}} t +\pi(2k+1)}{2}\right)}{\cosh^3 \left(\frac{e^{\frac{i\pi}{4}} t +\pi(2k+1)}{2}\right)} + (i \to -i ) \,. 
\end{equation}
This is an exact example whose singularities are consistent with \eqref{eq:resn}.\\
\indent Let us conclude with a few remarks.
\begin{itemize}
    \item Only the external singularities on the lattice in Figure \ref{fig:lattice1} have a clear bulk interpretation in terms of null geodesics, as they appear precisely when the two operators are separated by $\Delta t_{n}$ as in \eqref{eq:GeometricSection}.
    \item The transseries represention (\ref{capprox}) keeps exponentially small terms at large frequency. Since the result (\ref{eq:g12gd}) is valid uniformly in the complex plane as $|\omega|\to \infty$, it is plausible that it correctly captures all exponentially small corrections at large real frequency, but we do not have a proof of this statement. A useful consistency check of our framework would be to generalize the WKB approach of \cite{Afkhami-Jeddi:2025wra} to the other singularities beyond $t_{10}$. 
    \item The QNM asymptotics contain subleading power law corrections, which we neglected above. A correction to $\omega_n$ of the form $n^{-\gamma}$ leads to a subleading term $(t-t_{nm})^{d-2\Delta-1+\gamma}$ in the expansion around the singularities. For the $\text{AdS}_5$ black brane we have $\gamma=4/3$, and this matches the subleading correction found in \cite{Afkhami-Jeddi:2025wra}. 
\end{itemize}

\subsection{Imaginary QNMs and charged black holes}\label{chargedsec}
So far we have discussed the case where there is one line of asymptotic QNMs. Let us now generalize the analysis to the case of multiple asymptotic lines. The simplest example is a single line of imaginary poles, in addition to the line of complex poles (\ref{eq:qnmParameter}) considered previously. We take the asymptotic ansatz
\begin{align}
\omega_n=ian+ib+\ldots\hspace{10 mm}n\gg 1.
\end{align}
Applying the product formula (\ref{eq:productformula}) to this case gives 
\begin{align}\label{ccharged}
C(\omega)\approx \prod_\pm \Gamma\left(\frac{\pm i\omega+b}{a}+1\right)\Gamma\left(\frac{e^{i\theta}}{r}\left( \pm \omega+e^{-i\phi}s\right)+1\right) \Gamma\left(\frac{e^{-i\theta}}{r}\left( \pm \omega+e^{i\phi}s\right)+1\right).
\end{align}
In the presence of the extra line, the relation \eqref{eq:betaDelta} between the inverse temperature and the asymptotics of QNMs gets modified as follows \cite{Dodelson:2023vrw},
\be\label{eq:newbeta}
\beta = \frac{4\pi\sin \theta}{r}+\frac{2\pi}{a}. 
\ee
\begin{figure}[t]
\centering
\begin{tikzpicture}[scale=0.25]

  \pgfmathsetmacro{\piVal}{3.1416}
  \pgfmathsetmacro{\sqtwo}{1.4142}
  \pgfmathsetmacro{\offsetIm}{\piVal*(\sqtwo + 2.15)/2}  
  \pgfmathsetmacro{\scale}{\piVal/\sqtwo}        

    \fill[color=gray!20] (-20,-\offsetIm) rectangle (20,\offsetIm);
     \filldraw[ultra thick, black] (0,\offsetIm) circle (8pt) node[below right] {$\frac{i\beta}{2}$};
    \filldraw[ultra thick, black] (0,-\offsetIm) circle (8pt) node[above right] {$-\frac{i\beta}{2}$};
  \draw[dashed,gray, thick] (-20,-\offsetIm) -- (20,-\offsetIm);
    \draw[dashed,gray, thick] (-20,\offsetIm) -- (20,\offsetIm);
  \newdimen\imagcutoff
  \imagcutoff=20pt

  \draw[->] (-20,0) -- (20,0) node[right] {$\text{Re }t$};
  \draw[->] (0,-5) -- (0,21) node[above] {$\text{Im }t$};

\draw[dashed,brown, thick] (-20,\offsetIm+\offsetIm-\scale) -- (20,\offsetIm+\offsetIm-\scale);
\draw[dashed,brown, thick] (-20,\offsetIm+2*\offsetIm-2*\scale) -- (20,\offsetIm+2*\offsetIm-2*\scale);
\draw[dashed,brown, thick] (-20,\offsetIm+3*\offsetIm-3*\scale) -- (20,\offsetIm+3*\offsetIm-3*\scale);
\draw[dashed,brown, thick] (-20,\offsetIm+4*\offsetIm-4*\scale) -- (20,\offsetIm+4*\offsetIm-4*\scale);
\draw[dashed,gray, thick] (-20,2*\offsetIm) -- (20,2*\offsetIm);
\draw[dashed,gray, thick] (-20,3*\offsetIm) -- (20,3*\offsetIm);
  
  \foreach \m in {0,1,2,3,4,6,7,8,9,10} {
    \foreach \n in {0,1,2,3,4,6,7,8,9,10} {
      \foreach \k in {0,1,2,3,4,6,7,8,9,10} {
        \pgfmathparse{int(\m+\n+\k)}
        \let\sumint\pgfmathresult
        \ifnum\sumint<10
          \pgfmathsetmacro{\x}{(\offsetIm-\scale)*(\n - \m)}
          \pgfmathsetmacro{\y}{\offsetIm + (\offsetIm-\scale)*(\n + \m) + (2*\scale)*\k}
          \pgfmathsetlengthmacro{\ytmp}{\y pt}
          \ifdim \ytmp<\imagcutoff
            \filldraw[ultra thick, black] ({\x},{\y}) circle (8pt);
          \fi
        \fi
      }
    }
  }

  \coordinate (base) at (0,{\offsetIm});

  \pgfmathsetmacro{\xone}{(\offsetIm-\scale)*(1 - 0)}
  \pgfmathsetmacro{\yone}{(\offsetIm-\scale)*(1 + 0)}
  \draw[->,red,ultra thick] (base) -- ++({\xone},{\yone});

  \pgfmathsetmacro{\xtwo}{(\offsetIm-\scale)*(0 - 1)}
  \pgfmathsetmacro{\ytwo}{(\offsetIm-\scale)*(0 + 1)}
  \draw[->,red,ultra thick] (base) -- ++({\xtwo},{\ytwo});

  \pgfmathsetmacro{\xthree}{0}
  \pgfmathsetmacro{\ythree}{2*\scale}
  \draw[->,red,ultra thick] (base) -- ++({\xthree},{\ythree});
    \node at (\offsetIm-\scale+3,\offsetIm+\scale) {$v_+$};
  \node at (-\offsetIm+\scale-1,\offsetIm+\scale) {$v_-$};
  \node at (2,\offsetIm+2*\scale) {$v_a$};
\end{tikzpicture}
\caption{Lattice of singular points of $C(t)$ in the presence of a purely imaginary line of QNMs, in addition to the complex line considered previously. All singularities fall on the real sections $\text{Im }t=n\beta/2+m\beta_{\text{in}}/2$. To avoid clutter we only plot the real sections $\text{Im }t=n\beta/2$ and $\text{Im }t=\beta/2+n(\beta-\beta_{\text{in}})/2$, shown in dashed gray and dashed brown respectively.}
\label{fig:chlatt}
\end{figure}
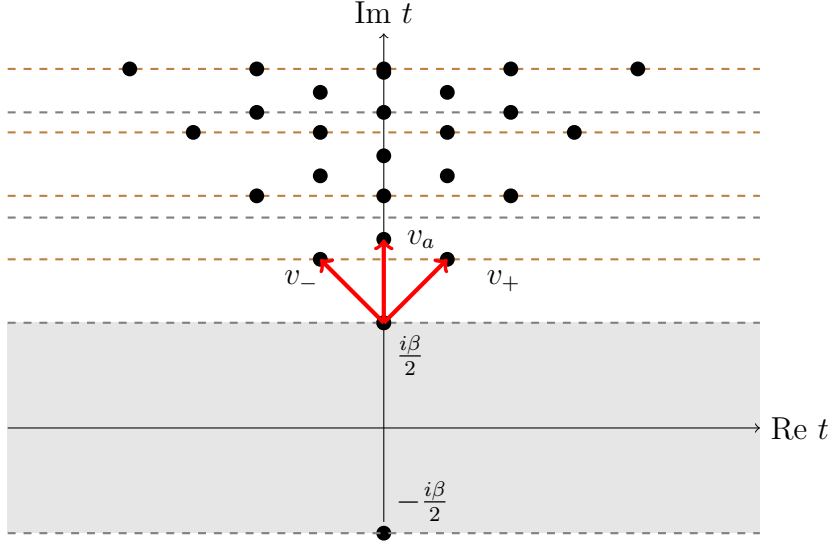
\indent The computation of the correlator in the time domain proceeds analogously to above. The only technicality is that the argument of the extra gamma functions in (\ref{ccharged}) is on a Stokes line. In this case we can use the following transseries representation \cite{Nemes_2015}
\begin{align}\label{transseries}
\prod_\pm \Gamma\left(\frac{\pm i\omega+b}{a}+1\right)\approx \frac{\pi}{\sinh\left(\frac{\pi \omega}{a}\right)}\left(\frac{\omega}{a}\right)^{1+\frac{2b}{a}}.
\end{align}
Readers who are skeptical of transseries are encouraged to set $b=0$,  for which (\ref{transseries}) becomes an identity. Performing the Fourier transform as above, we find singularities on a lattice with three basis vectors, 
\begin{align}
t_{nmk}=\frac{i\beta}{2}+nv_++mv_-+kv_{a},\hspace{10 mm}n,m,k\in \mathbb{Z}_{\ge 0}.
\end{align}
The new basis vector is given by $v_a=2\pi i/a$, and the resulting lattice is depicted in Figure \ref{fig:chlatt}. The imaginary part of the singularities is given by 
\begin{align}\label{imtcharged}
\text{Im }t_{nmk}=\frac{\beta}{2}(1+n+m)+\frac{\pi(2k-m-n)}{a}.
\end{align}
Note that $\beta/2>\pi/a$ as a result of \eqref{eq:newbeta}, which ensures that all singularities satisfy $|\text{Im }t_{mnk}|>\beta/2$ as expected. \\
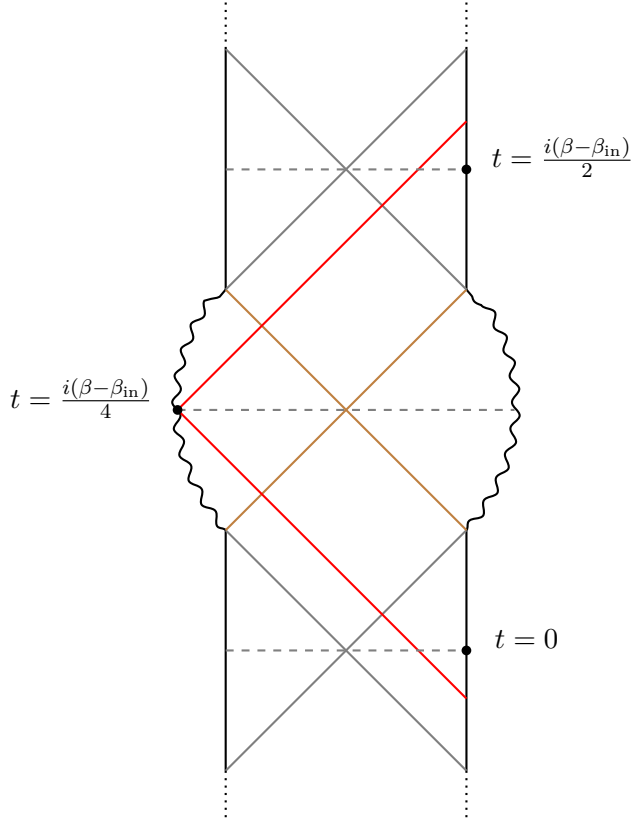
\begin{figure}[t]
\centering
\begin{tikzpicture}[scale=1.6, thick]
\coordinate (Ltop) at (-1,1);
\coordinate (Lbot) at (-1,-1);
\coordinate (Rtop) at (1,1);
\coordinate (Rbot) at (1,-1);
\coordinate (LStop) at (-1,3);
\coordinate (RStop) at (1,3);
\coordinate (LSbtop) at (-1,5);
\coordinate (RSbtop) at (1,5);
\draw[black] (Ltop) -- (Lbot);
\draw[black] (Rtop) -- (Rbot);
\draw[gray] (Lbot) -- (Rtop);
\draw[gray] (Ltop) -- (Rbot);
\draw[black, decorate, decoration={snake, amplitude=1.5pt}] (Ltop) to[out=130, in=230] (LStop);
\draw[black, decorate, decoration={snake, amplitude=1.5pt}] (Rtop) to[out=50, in=-50] (RStop);
\draw[brown] (Ltop) -- (RStop);
\draw[brown] (Rtop) -- (LStop);

\draw[black] (LStop) -- (LSbtop);
\draw[black] (RStop) -- (RSbtop);

\draw[gray] (LStop) -- (RSbtop);
\draw[gray] (RStop) -- (LSbtop);

\draw[dotted] (RSbtop) -- (1,5.4);
\draw[dotted] (LSbtop) -- (-1,5.4);

\draw[dotted] (Rbot) -- (1,-1.4);
\draw[dotted] (Lbot) -- (-1,-1.4);

\draw[red, thick] (1,-0.4)--(-1.4,2);
\draw[red, thick] (-1.4,2)--(1,4.4);

\draw[gray, dashed] (-1,0) -- (1,0);
\draw[gray, dashed] (-1.4,2) -- (1.4,2);
\draw[gray, dashed] (-1,4) -- (1,4);

\node[circle, fill, inner sep=1.3pt] at (1,0) {};
\node at (1.5,0.1) {\small $t=0$};
\node[circle, fill, inner sep=1.3pt] at (-1.4,2) {};
\node at (-2.2,2.1) {\small $t=\frac{i(\beta-\beta_{\text{in}})}{4}$};
\node[circle, fill, inner sep=1.3pt] at (1,4) {};
\node at (1.8,4.1) {\small $t=\frac{i(\beta-\beta_{\text{in}})}{2}$};
\end{tikzpicture}
\caption{A bouncing geodesic in the maximally extended Reissner-Nordström-AdS black brane. Similarly to what happens in the uncharged case, in order for the geodesic to meet the singularity at $\text{Re }t=0$, it must start with boundary time $t<0$. In contrast to the case of a spacelike singularity, every bouncing geodesic that connects two boundaries must cross an inner horizon (shown in brown).}
\label{fig:chpenbg}
\end{figure}
\indent From (\ref{imtcharged}), we notice that some of the singularities have moved off the real sections at $\text{Im }t=n\beta/2$. The new locations of the singularities can be understood in the bulk by considering the example of a charged black brane in AdS$_{5}$ \cite{Brecher:2004gn,Balasubramanian:2019qwk}, which has a line of imaginary poles \cite{Jansen:2017oag}. The metric is given by
\be
ds^2 = -f(r) dt^2+\frac{dr^2}{f(r)}+r^2 d\Omega^2_3 \,, \hspace{10 mm} f(r) = r^2-\frac{\mu}{r^2}+\frac{Q^2}{r^4} \,.
\ee
The geometry has an outer and an inner horizon, which we denote by $r_+$ and $r_-$ respectively. The two associated temperatures are given by
\begin{equation}
    \beta = \frac{4\pi}{f'(r_+)},\hspace{10 mm} \beta_{\text{in}}=  -\frac{4\pi}{f'(r_-)} \,.
\end{equation}
For generic values of $Q$, one has $\beta \ge\beta_{\text{in}}$.

\indent Similarly to the uncharged case, the singularity in charged black branes should be represented as bending outward in the Penrose diagram \cite{Brecher:2004gn,Balasubramanian:2019qwk}. Indeed, null geodesics starting from different boundaries at zero real boundary time separation do not meet at the singularity. The situation is depicted in Figure \ref{fig:chpenbg}, with the red geodesic connecting two boundary points separated  by complex time $\Delta t$ with 
\be
\text{Im }\Delta t = 2 \int_0^{\infty} \frac{dr}{f(r)} = \frac{\beta-\beta_{\text{in}}}{2} \,,
\ee 
again with a proper choice of $i\epsilon$ prescription. Note that the relative minus sign comes from the fact that $f'(r_-)<0$.  More generally, a geodesic bouncing around the Penrose diagram of the charged black brane $n$ times will connect points separated by complex time $\Delta t_n$, with
\be\label{chargedbounce}
\text{Im } \Delta t_n = \frac{n(\beta-\beta_{\text{in}})}{2},\hspace{10 mm} n \in \mathbb{Z}_{>0} \,.
\ee
We remark that as a consequence of the fact that the singularity is now timelike, $\text{Im }\Delta t_n$ depends on the inner temperature, and is therefore shifted off the real sections at $\text{Im }t=n\beta/2$. However, as shown in Figure 4, the addition of a charge introduces new real sections that are obtained by crossing the inner horizon a given number of times. The most general real section takes the form $\text{Im }t=n\beta/2+m\beta_{\text{in}}/2$ with $m,n\in\mathbb{Z}$, so that (\ref{chargedbounce}) indeed lies on a real section.  \\
\indent We now compare this bulk picture with the previous result (\ref{imtcharged}). The imaginary part of the singularities at $t_{n00}$ and $t_{0m0}$ matches the expression (\ref{chargedbounce}) for a bouncing geodesic as long as the spacing between imaginary QNMs satisfies
\begin{align}\label{abetain}
a=\frac{2\pi}{\beta_{\text{in}}}.
\end{align}
We checked this relation numerically by computing the QNMs using $\mathtt{QNMSpectral}$ \cite{Jansen:2017oag} and found agreement. It would be interesting to find an independent analytic proof of (\ref{abetain}). Analogously to the uncharged case, the outermost singularities in the lattice all have a geometric interpretation in term of bouncing geodesics, and all the singularities (\ref{imtcharged}) lie on real sections.

Let us conclude this section by briefly discussing the generic case of $K$ asymptotic lines of QNMs. We assume them to be at generic angles, that is
\be
\omega_{ni} = r_i e^{i\theta_i} n + s_i e^{i \phi_i} + \dots
\ee
The relation between the asymptotic parameters and the inverse temperature now becomes \cite{Dodelson:2023vrw}
\be
\beta = \sum_{i=1}^K \frac{4\pi \sin \theta_i}{r_i} \,.
\ee
The lattice of singularities of $C(t)$ takes the form
\be
t_{\vec{n}\vec{m}} = \frac{i \beta}{2}+ \left(\vec{n}\cdot \vec{v}_{+}+ \vec{m}\cdot \vec{v}_- \right), \hspace{10 mm} v_{\pm i}= \pm \frac{2\pi}{r_i} e^{\pm i \theta_i} .
\ee
Assuming $0<\theta_1<\theta_i<\pi/2$ for $i=2,\dots,K$, the outermost singularities are  $(n_i,m_i)=(\delta_{1i},0)$ and $(n_i,m_i)=(0,\delta_{1i})$. For generic values of the parameters these points are all off the real sections $\text{Im }t=n\beta/2$, in complete analogy with the charged case. The presence of different branches of QNMs at generic angles was related to higher derivative corrections to the bulk metric in \cite{Grozdanov:2016vgg,Grozdanov:2018gfx}.

\section{The effect of zeroes}\label{sec:zeroes}
In the previous section, we found that the singularities of $C(t)$ in complex time form a lattice in the case of holographic theories. This structure is universal for any asymptotically AdS black hole. It is natural to ask what happens to the lattice at finite coupling, away from the maximally chaotic limit.   \\
    \indent The key properties that lead to the lattice in Section \ref{sec:Holography} were meromorphy of the correlator and the absence of zeroes in the complex frequency plane. At finite 't Hooft coupling, there is good reason to believe that meromorphy is preserved, at least for a specific class of models. Indeed, classical systems that are sufficiently chaotic have meromorphic correlators \cite{Pollicott1985,PhysRevLett.56.405}, and in \cite{Prosen_2002,Dodelson:2024atp,Dodelson:2025rng} this property was studied for quantum systems including SYK. The interpretation of zeroes in frequency space is much less clear, and in fact it was shown in \cite{Dodelson:2024atp} that the SYK correlator at high temperatures does possess zeroes (see Appendix \ref{qnmsyk} for an improved computation). This suggests that we investigate the effect of zeroes on the lattice structure.

\begin{figure}[t]
\centering
\begin{tikzpicture}[scale=0.5]

    \fill[color=gray!20] (-10,-pi+1/2) rectangle (10,pi-1/2);
  \draw[dashed,gray, thick] (-10,pi-1/2) -- (10,pi-1/2);
   \draw[dashed,gray, thick] (-10,-pi+1/2) -- (10,-pi+1/2);
    \draw[dashed,gray, thick] (-10,2*pi-1) -- (10,2*pi-1);
    \draw[dashed,gray, thick] (-10,3*pi-3/2) -- (10,3*pi-3/2); 
    
  \draw[->] (-10,0) -- (10,0) node[right] {$\text{Re}\,t$};
  \draw[->] (0,-pi+1/2) -- (0,10) node[above] {$\text{Im}\,t$};

  \coordinate (z0) at (0,pi-1/2);
  \coordinate (z1) at (pi,2*pi-1/2);
  \coordinate (z11) at (-pi,2*pi-1/2);
  \coordinate (z2) at (2*pi,3*pi-1/2);
  \coordinate (z22) at (-2*pi,3*pi-1/2);
  \coordinate (z222) at (0,3*pi-1/2);

    \filldraw[black] (z0) circle (5pt) node[below right, black] {$\frac{i\beta}{2}$};
    \filldraw[black] (z0)++(0,1) circle (5pt);
    \filldraw[black] (0,-pi+1/2) circle (5pt) node[above right, black] {$-\frac{i\beta}{2}$};
    \filldraw[black] (z1) circle (5pt);
    \filldraw[black] (z1)++(0,1) circle (5pt);
    \filldraw[black] (z11) circle (5pt);
    \filldraw[black] (z11)++(0,1) circle (5pt);
    \filldraw[black] (z2) circle (5pt);
    \filldraw[black] (z2)++(0,1) circle (5pt);
    \filldraw[black] (z22) circle (5pt);
    \filldraw[black] (z22)++(0,1) circle (5pt);
    \filldraw[black] (z222) circle (5pt);
    \filldraw[black] (z222)++(0,1) circle (5pt);

  \draw[->, ultra thick, red] (z0) -- (z1);
  \draw[->, ultra thick, red] (z0) -- (z11);
  \draw[|->, ultra thick, blue] (2*pi-0.75,3*pi-1/2) -- (2*pi-0.75,3*pi+1/2);
  \node at (2*pi-1.65,3*pi+0.1) {$\alpha'$};

  \node at (3,4.3) {$v_+$};
  \node at (-3.4,4.3) {$v_-$};

\end{tikzpicture}
\caption{Singularities of $C(t)$ for the toy model (\ref{eq:ProductWZeroes}) with a line of zeroes. Each singularity splits into two points separated by a distance $\alpha'$ in the presence of the zeroes. The real sections $\text{Im }t=n\beta/2$ are depicted as dashed lines.}
\label{fig:latticezeroes}
\end{figure}
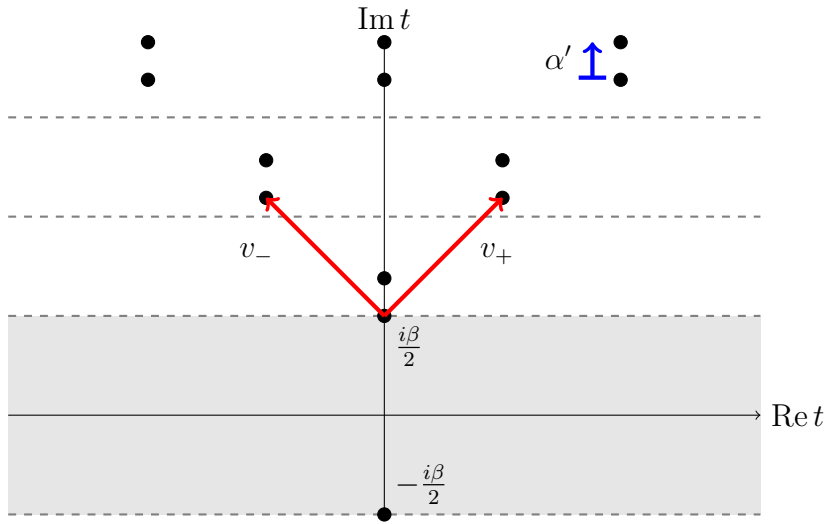

\indent Let us therefore consider what happens when we introduce zeroes in the complex frequency plane. A finite number of zeroes will not change the location of the bouncing singularities, but an infinite number will have an effect. For instance, let us start with a single line of QNMs of the form (\ref{eq:qnmParameter}) and add an infinite family of zeroes along the imaginary axis,
\be\label{eq:ProductWZeroes}
C(\omega) = \frac{\cosh(\frac{\alpha'\omega}{2})}{\prod_{n=1}^\infty \left(1-\frac{\omega^2}{\omega_n^2}\right)\left(1-\frac{\omega^2}{(\omega_n^*)^2}\right)}.
\ee 
Matching to the asymptotic behavior $C(\omega)\sim e^{-\beta \omega/2}$ at large positive frequency \cite{Caron-Huot:2009ypo} gives
\begin{align}\label{modified}
\beta=\frac{4\pi \sin\theta}{r}-\alpha',
\end{align}
which modifies the relation (\ref{eq:betaDelta}) between the spacing of QNMs and the inverse temperature.  More generally, we could consider complex lines of zeroes preserving the invariance of $C(\omega)$ under $\omega\to-\omega$ and $\omega\to\omega^*$, but an imaginary line of zeroes will suffice to illustrate the point.\\
\indent The new line of zeroes has two important effects: it shifts the singularities off the real sections at $\text{Im }t=n\beta/2$, and destroys the lattice structure. To see this, we repeat the calculation of Section \ref{sec:Holography} for the correlator (\ref{eq:ProductWZeroes}).
As shown in Figure \ref{fig:latticezeroes}, we find singularities of the form
\be\label{eq:singZeroes}
t_{nmj} = \frac{i\beta}{2}+ n v_+ +m v_-+i\alpha'\delta_{j1},\hspace{10 mm}n,m\in \mathbb{Z}_{\ge 0},\hspace{10 mm}j=0,1.
\ee 
Here $v_{\pm}=\pm 2\pi e^{\pm i\theta}/r$ as before. We see that the Kronecker delta term ruins the additivity of the lattice. In addition, using the modified relation (\ref{modified}), we find the shifted imaginary part
\be 
\text{Im }t_{nmj}= \frac{\beta}{2}(1+n+m)+\frac{\alpha'}{2}(n+m+2\delta_{j1}).
\ee 
For generic $\alpha'$, this is only on a real section when $n=m=j=0$. \\
\indent Note that keeping $\alpha'$ finite is necessary in order to shift the singularities. At $k$th order in $\alpha'$, the correlator (\ref{eq:ProductWZeroes}) behaves as $(\alpha' w)^k$ times the zeroth order answer. In position space, this leads to singularities on the real sections that become stronger at higher orders in perturbation theory. It is only after resumming all orders of perturbation theory that the singularities are shifted.

\section{Singularity resolution at finite coupling}\label{sec:SYK}
In this section, we present a microscopic example of the singularity structure of $C(t)$ in complex time at finite coupling. We will see that the singularities move off the real sections into the complex plane. The heuristic interpretation is that the bouncing geodesic singularities are smoothed out into finite-size bumps by stringy effects. This is reminiscent of bulk cone singularities in one-sided correlators \cite{Hubeny:2006yu,Dodelson:2023nnr}, where the relevant stringy effects are tidal forces \cite{Dodelson:2020lal}. Note that the bulk point singularity in vacuum correlators is also resolved in string theory, but in that case the singularity is entirely removed from the complex plane \cite{Maldacena:2015iua}. This is an important qualitative difference from the correlator we study here, where stringy effects do not remove the singularity completely. \\
\indent We will work with the SYK model \cite{Maldacena:2016hyu,kitaev,Sachdev_1993,Polchinski:2016xgd}, which describes $N$ Majorana fermions interacting via a random Gaussian coupling. The Hamiltonian is 
\begin{align}
H=i^{q/2}\sum_{1\le i_1\le\ldots \le i_q\le N}j_{i_1\ldots i_q}\psi_{i_1}\cdots \psi_{i_q},\hspace{10 mm}\langle j^2_{i_1\ldots i_q}\rangle=\frac{(q-1)!\mathcal{J}^2}{2^{1-q}qN^{q-1}}.
\end{align}
The correlator of interest is the two-point function of elementary fermions at infinite $N$ in the infinite temperature limit,
\begin{align}
C(t)=2\text{Tr}(\psi_1(t)\psi_1(0)).
\end{align}
\indent The motivation for studying this model is two-fold. First, since we are interested in analytically continuing the two-point function into the complex plane, we need to be able to compute correlators to a high level of accuracy. This is possible in the SYK model at infinite $N$, due to the Schwinger-Dyson equation satisfied by the two-point function. At infinite temperature, this equation can be efficiently solved perturbatively in the coupling $\mathcal{J}$, leading to the moment expansion \cite{Parker:2018yvk,Dodelson:2024atp}
\begin{align}\label{momentexp}
C(t)=\sum_{n=0}^{\infty}\frac{\mu_{2n}}{(2n)!}(it)^{2n},\hspace{10 mm}\mu_{2n}\propto \mathcal{J}^{2n}.
\end{align}
The moments $\mu_{2n}$ were computed up to $n=2000$ in \cite{Dodelson:2024atp}, and this turns out to be sufficient precision for our purposes.  \\
\indent The second reason to study SYK is that the model is in a certain sense very close to maximally chaotic. At low temperatures, the system is holographic and saturates the chaos bound \cite{Maldacena:2015waa}. But even at infinite temperature, a related bound is almost saturated. Indeed, recall that the asymptotic behavior of the moments as $n\to \infty$ is controlled by the universal operator growth hypothesis \cite{Parker:2018yvk},
\begin{align}
\mu_{2n}\sim\left(\frac{4n}{e\beta_0}\right)^{2n},\hspace{10 mm}n\to \infty.
\end{align}
The parameter $\beta_0$ is analogous to a temperature, and provides an upper bound for the Lyapunov exponent $\lambda$ \cite{Parker:2018yvk},
\begin{align}\label{chaosbound}
\lambda<\frac{2\pi}{\beta_0}.
\end{align}
This is a generalization of the ordinary chaos bound to systems with no coincident point singularity in the two-point function. In $q=4$ SYK, the left hand side of (\ref{chaosbound}) is 1.240, and the right hand side is 1.246 \cite{Parker:2018yvk}. Here we are working in units where $\mathcal{J}=1$. For larger $q$, the bound is even closer to being saturated. Since stringy corrections lead to a submaximal Lyapunov exponent \cite{Shenker:2014cwa}, we might expect the system to behave like an AdS black hole with small but finite $\alpha'$, so that the singularities are only slightly off the real sections defined by $\text{Im }t=n\beta_0/2$. We will see shortly that this intuition indeed holds.
\subsection{Singularities for various $q$}
\begin{figure}
        \centering
        \includegraphics[width=.6\textwidth]{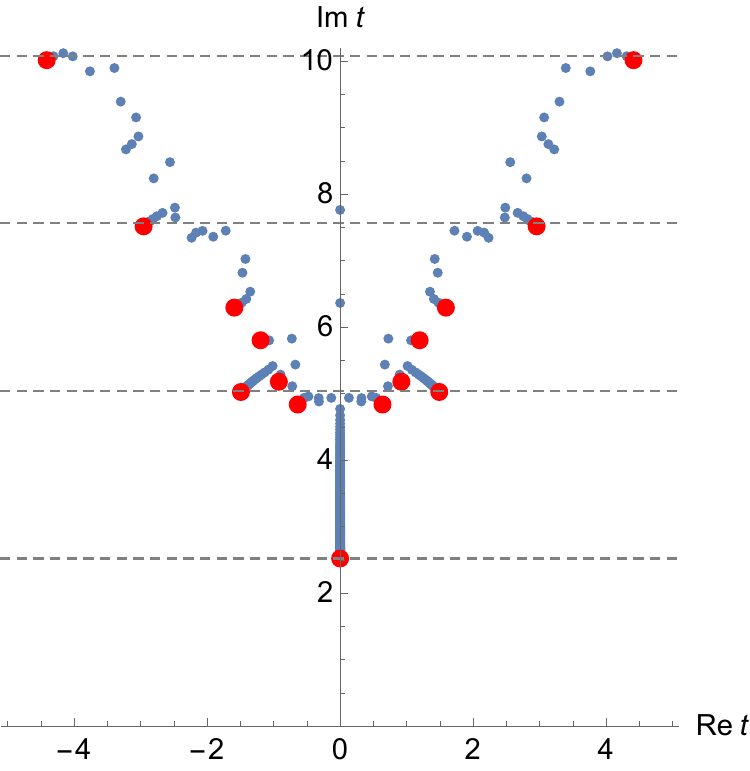}
        \caption{The structure of singularities for $q=4$ SYK. The converged poles of the $(1000,1000)$ Padé approximant are shown in red, and lie at the endpoints of branch cuts formed by accumulating poles. The outermost red dots are slightly off the real sections $\text{Im }t=n\beta_0/2$, which are displayed as dashed lines. \label{polesqeq4}}
\end{figure}
\begin{figure}[t]
    \centering
    \begin{subfigure}[b]{0.32\textwidth}
        \centering
             \includegraphics[width=\textwidth]{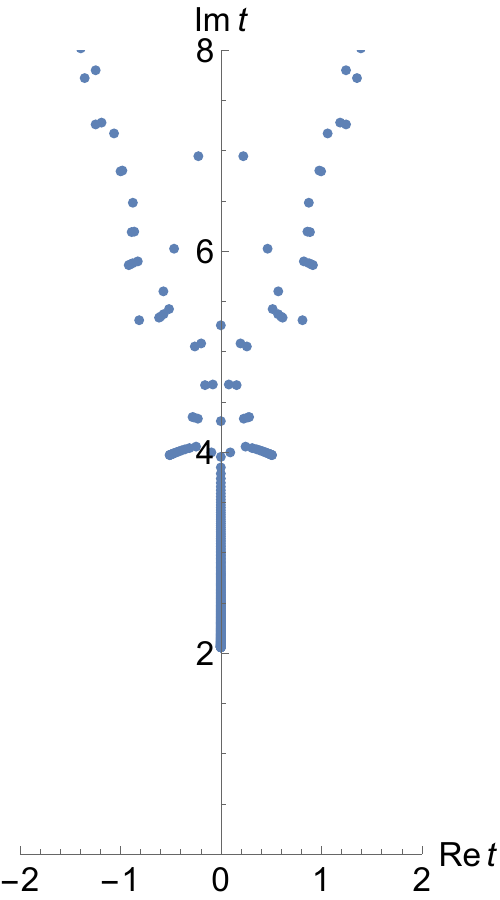}
        \caption{}
    \end{subfigure}
    \hfill
    \begin{subfigure}[b]{0.32\textwidth}
        \centering
             \includegraphics[width=\textwidth]{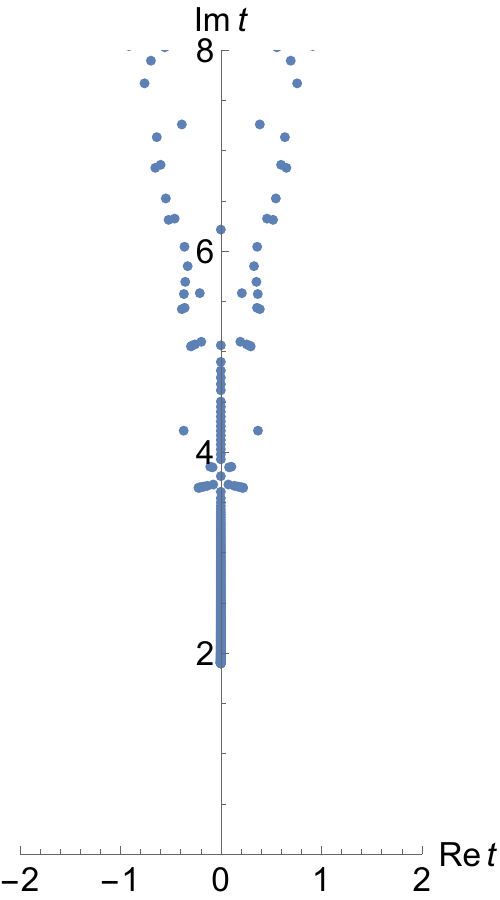}
        \caption{}
    \end{subfigure}
    \hfill
    \begin{subfigure}[b]{0.32\textwidth}
        \centering
           \includegraphics[width=\textwidth]{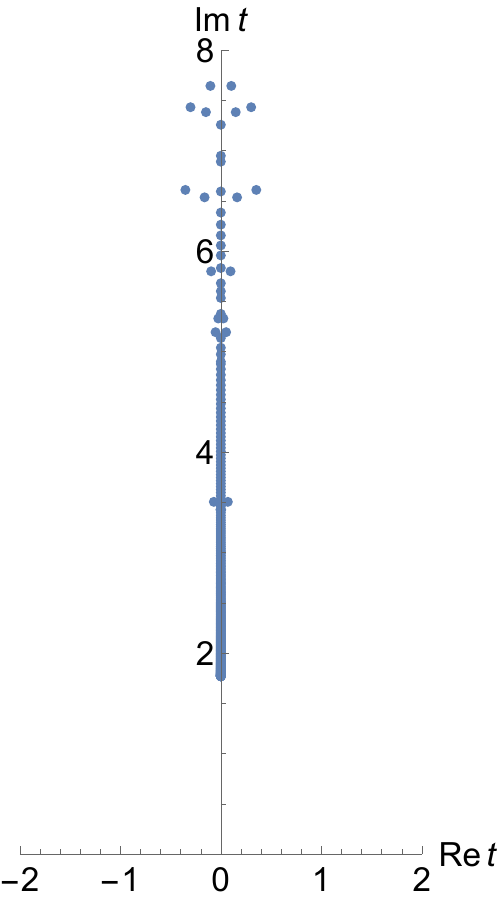}
        \caption{}
    \end{subfigure}
    \caption{The complex $t$ plane for (a) $q=6$, (b) $q=8$, and (c) $q=12$. As $q$ increases, the singularities collapse onto the imaginary axis. Here the $(750,750)$ Pad\'e approximations are shown. \label{qlarge}}
\end{figure}
The moment expansion (\ref{momentexp}) defines the correlator within the radius of convergence $|t|<\beta_0/2$. In order to analytically continue $C(t)$ outside this range, we employ the technique of Padé approximation, which approximates a power series by the ``best possible" rational function. For simplicity, we choose the Padé approximant to have the same number of poles and zeroes. The number of poles should then be taken to be half of the number of available moments. \\
\indent Let us start with the case $q=4$, for which 2000 moments have been computed. The poles of the Padé approximant of order 1000 are depicted as blue dots in Figure \ref{polesqeq4}. In red, we have displayed the poles that have converged within tolerance $10^{-6}$, by which we mean that the Padé approximants of order 900 and 1000 contain poles that are separated by a distance less than $10^{-6}$. In order to compute the leading singularity off the imaginary axis with an accuracy of $10^{-2}$, one needs to input 45 moments. We note that there is an accumulation of poles starting at the red points. This is a signature of branch cut formation \cite{STAHL1997139}.\\
\indent As is evident from Figure \ref{polesqeq4}, the outermost singularities almost lie on the real sections at $\text{Im }t=n\beta_0/2$. For example, there is a singularity at $t=\pm 1.49+5.02i$, which is a distance $.02$ from the real section at $\text{Im }t=\beta_0=5.04$. There are also singularities close to the higher real sections, confirming the expectation that this system behaves like an AdS black hole with small string scale. As expected from the previous section, the singularities are not closed under addition, and therefore do not form a lattice. \\
\indent We can also repeat the computation for higher values of $q$, for which we have data up to 1500 moments. In the infinite $q$ limit, we should reproduce the answer for AdS$_2$ \cite{Tarnopolsky:2018env,Lin:2023trc}, which does not have a black hole singularity. The bouncing geodesic singularities should therefore disappear in this limit. As $q$ is increased, we indeed find that the singularities move toward the imaginary axis, as shown in Figure \ref{qlarge}. For $q=20$, all the singularities are on the imaginary axis.\\\indent The methods that we have described here are effective for obtaining the singularities towards the outside of the tree in Figure \ref{polesqeq4}, but we expect more singularities to appear when more moments are taken into account. Although it is difficult to make any precise statement with our limited data, it is tempting to speculate on the analytic structure of $C(t)$ deep into the complex plane. The pattern in Figure \ref{polesqeq4} suggests that the density of singularities grows as one probes further into the complex plane. One possibility is that a natural boundary forms along some curve, beyond which the correlator cannot be analytically continued. This is what generically happens for solutions of the equations of motion in classical chaotic theories, whose singularities reflect the presence of intermittent bursts in real time \cite{PhysRevA.23.2673}. In that setup, the natural boundary has fractal dimension greater than one \cite{CHANG1981211,chang1982analytic}. It would be interesting to understand whether the same is true for the SYK correlator.

\subsection{The power of the singularity}
So far we have analyzed the locations of the singularities in the complex plane. Let us now discuss their functional form. After using Padé approximation to obtain an analytic continuation of $C(t)$, we can numerically fit the correlator near a given singularity to a power law behavior. For fixed $q$, we find that all the converged branch point singularities at $t=t_n$ have the same power,
\begin{align}\label{ctaupower}
C(t)\sim \frac{a(t_n)}{(t-t_n)^{2\Delta}},\hspace{10 mm}\Delta=\frac{1}{q-2},\hspace{10 mm}q>4.
\end{align}
This is a branch cut singularity. For $q=4$ the cut becomes logarithmic,
\begin{align}
C(t)\sim \frac{a(t_n)}{t-t_n}+b(t_n)\log(t-t_n),\hspace{10 mm}q=4.
\end{align}
\indent We can check the result (\ref{ctaupower}) in the large $q$ limit, where the correlator can be studied analytically. The large $q$ expansion is given by \cite{Bhattacharjee:2022ave}
\begin{align}
C(i\tau)=1-\frac{2}{q}\log \cos\tau+\frac{1}{q^2}h(\tau)+O\left(\frac{1}{q^3}\right),
\end{align}
where we have defined
\begin{align}
h(\tau)&=2(\log \cos\tau)^2-2\ell(\tau)+4\tan\tau\int_{0}^{-\tau}dy\, \ell(y)+4\tau\tan\tau\\
\ell(\tau)&=-2\log \cos\tau-\cos^2 \tau\text{Li}_2(-\tan^2\tau).
\end{align}
The result of the integral is somewhat lengthy, and the full expression can be found in \cite{Bhattacharjee:2022ave}. Expanding around $\tau=\pi/2$, we obtain
\begin{align}\label{largeqsing}
C(i\tau)&\sim 1-\frac{2}{q}\log\left(\frac{\pi}{2}-\tau\right)\notag\\
&\hspace{10 mm}+\frac{1}{q^2}\left[\frac{\pi(6-\pi^2)}{3\left(\frac{\pi}{2}-\tau\right)}+2\log\left(\frac{\pi}{2}-\tau\right)^2-4\log\left(\frac{\pi}{2}-\tau\right)\right],\hspace{10 mm}\tau\sim \frac{\pi}{2}.
\end{align}
\indent In order to match (\ref{largeqsing}) with the numerics, we make the following ansatz for the behavior near $\tau=\beta_0/2$,
\begin{align}\label{poweransatz}
C(i\tau)\sim \frac{1}{\left(\frac{\beta_0}{2}-\tau\right)^{2\Delta}},\hspace{10 mm}\tau \sim \frac{\beta_0}{2}.
\end{align}
Comparing with (\ref{largeqsing}), we find
\begin{align}
\beta_0&=\pi+\frac{\pi(\pi^2-6)}{3q}+\ldots\\
\Delta&=\frac{1}{q}+\frac{2}{q^2}+\ldots
\end{align}
This indeed matches the large $q$ expansion of (\ref{ctaupower}). \\
\indent The scaling law $\Delta=1/(q-2)$ is somewhat unexpected. It does not match the infrared scaling dimension $1/q$ of $\psi$ \cite{Maldacena:2016hyu}. Nor does it reproduce the short distance scaling dimension (which is zero). In order to understand this result, let us consider the equation of motion, 
\begin{align}
\dot{\psi}_1=i^{q/2-1}\sum_{1<i_1<\ldots<i_{q-1}\le N}j_{1i_1\ldots i_{q-1}}\psi_{i_1}\cdots \psi_{i_{q-1}}.
\end{align}
The two sides of this equation are comparable at $t=t_n$ if $\psi_i\sim 1/(t-t_n)^{1/(q-2)}$. This explains the scaling dimension that we have found. Equivalently, the action is rescaled under a similarity transformation \cite{Biggs:2023sqw}, 
\begin{align}
\psi_i'( \lambda t)=\lambda^{-\frac{1}{q-2}}\psi_i( t),\hspace{10 mm}S'=\lambda^{-\frac{2}{q-2}}S.
\end{align}
This is not a symmetry of the theory, but leaves the equations of motion invariant. \\
\indent Since $\Delta$ is fractional, the correlator is defined on a complicated multi-sheeted Riemann surface. The same is true of solutions to the equations of motion in classical few-body chaotic systems \cite{chang1982analytic,CHANG1981211}. In fact, the multi-sheeted nature of classical solutions is sometimes taken to be a defining property of chaos, so that only integrable systems can have globally meromorphic solutions. This is known as the Painlevé property \cite{ramani1989painleve}. Here we have seen that this signature of chaos can be extended to correlators in quantum many-body systems. Of course, in a generic interacting field theory (such as $\mathcal{N}=4$ super Yang-Mills), the power $\Delta$ need not be rational. 
\section{Outlook}\label{outlook}
In this paper we have investigated signatures of the black hole singularity in thermal correlators at finite 't Hooft coupling. In the case of the SYK model, we found that the bouncing singularities are shifted slightly off the real sections. It is important to understand whether this behavior is a universal feature of chaotic large $N$ systems. For instance, the BFSS matrix model \cite{Banks:1996vh} exhibits bouncing singularities at strong coupling \cite{Biggs:2023sqw}, but their fate at finite coupling is unknown. The matrix bootstrap \cite{Lin:2020mme} along with classical simulations might be helpful for addressing this question. The eventual goal is to study gauge theories in higher dimensions, which will likely require a new set of tools.
\\ \indent One direction that we have not attempted to explore is quantum gravitational effects in the bulk. Stringy corrections only shift the bouncing singularities off the real sections; perhaps $1/N$ effects remove them entirely. Quantum gravity may therefore have an important role to play for the thermal two-point function, which is not the case for the bulk point singularity in a weakly coupled bulk \cite{Maldacena:2015iua}. Specific quantum effects that could be important include wormholes, instantons \cite{Kruthoff:2024gxc}, topology change, gravitational radiation, and self-force interactions. It would be interesting to identify which of these (if any) is responsible for removing the singularity. \\
\indent Our results suggest that certain features of black holes survive away from the maximally chaotic limit. In higher dimensions, black hole physics becomes much richer, with various phenomena such as exponentially long-lived QNMs \cite{Festuccia:2008zx,Berenstein:2020vlp,Dodelson:2022eiz}, accumulating modes at large spatial momentum 
\cite{Festuccia:2008zx}, and quasiparticles with superluminal dispersion that probe the photon sphere \cite{Dodelson:2023nnr}. How does this picture change at finite coupling? For instance, is there some notion of a stringy photon sphere? We hope that the higher dimensional melonic theories presented in \cite{Murugan_2017,Chang:2021fmd} are a setting in which such questions can be quantitatively addressed.

\section*{Acknowledgments}
We would like to thank Gauri Batra, Simon Caron-Huot, Nejc Čeplak, Yiming Chen, Daniel Jafferis, Adam Levine, Hong Liu, Juan Maldacena, Andrei Parnachev, Stephen Shenker, Ahmed Sheta,  Douglas Stanford, Andrew Strominger, Haitian Xie, and Alexander Zhiboedov for helpful input, and Stephen Shenker and Ahmed Sheta for comments on the manuscript. We are grateful to the Simons Center for Geometry and Physics for hosting and supporting the program ``Black Holes and Strongly Coupled Thermal Dynamics," as well as the program participants for useful discussions. This work was performed in part at the Aspen Center for Physics, which is supported by National Science Foundation grant PHY-2210452. MD's work is supported by DOE grant DE-SC/0007870 and the Frankel-Goldfield
Research Fund. CI is partially supported by the Swiss National Science Foundation Grant No.
185723 and the NCCR SwissMAP. R.K.\ is
supported by the Titchmarsh Research Fellowship at the Mathematical Institute and by the Walker Early Career Fellowship at Balliol College. RK has also received support from the European Research Council (ERC) under the European Union’s Horizon 2020 research and innovation program (grant agreement number 949077).  For the purpose of open access, the authors have applied a CC BY public copyright licence to
any Author Accepted Manuscript (AAM) version arising
from this submission.

\appendix
\section{Analyticity of thermal correlators}\label{app:ancont}

In this appendix we review the general analytic properties of the two-point function in the complex time plane \cite{Iliesiu:2018fao}\footnote{The analytic structure of finite temperature CFT correlators has also recently been explored in \cite{Alday:2020eua, Barrat:2025nvu,Barrat:2025twb,Buric:2025anb,Buric:2025fye}.}. We start with the definition
\begin{align}\label{cdef}
C(t,x)=\text{Tr}\left(e^{-\beta H}\mathcal{O}\left(t-\frac{i\beta}{2},\vec{x}\right)\mathcal{O}(0,0)\right),\hspace{10 mm}x\equiv |\vec{x}|.
\end{align}
This is analytic in the strip $-\beta/2<\text{Im }t<\beta/2$. In addition, it satisfies the KMS condition $C(-t,x)=C(t,x)$. On the edge of the strip of analyticity, there are branch points at $t=i\beta/2\pm x$ and $t=-i\beta/2\pm x$.\\
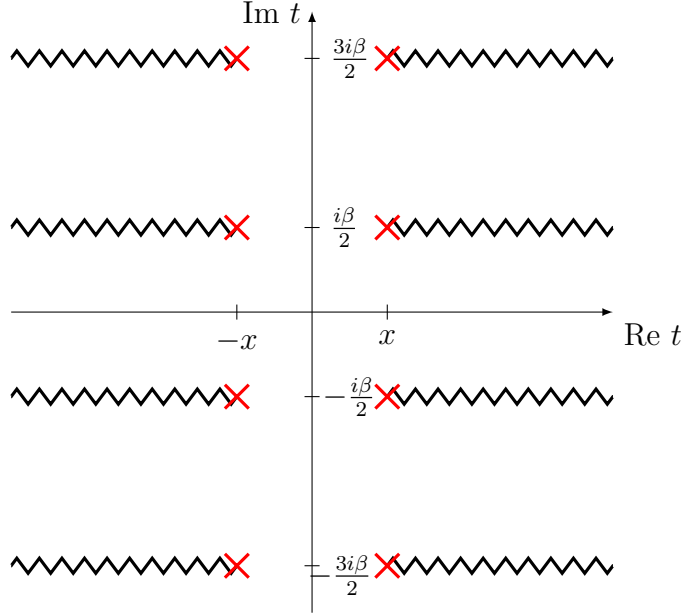
\begin{figure}
    \centering
    \begin{tikzpicture}[scale=1,>=latex]

\draw[->] (-4,0) -- (4,0) node[below right] {$\text{Re }t$};
\draw[->] (0,-4) -- (0,4) node[left] {$\text{Im }t$};

\def\period{1.5}
\def\ncross{2} 

\foreach \k in {-2,...,1} {
  \pgfmathsetmacro{\y}{\k*\period}

  \draw[very thick, black, decorate, decoration={zigzag, segment length=3mm, amplitude=1mm}] (1,1.5*\y+1.125) -- (4,1.5*\y+1.125);
  \draw[very thick, black, decorate, decoration={zigzag, segment length=3mm, amplitude=1mm}] (-1,1.5*\y+1.125) -- (-4,1.5*\y+1.125);

  \node[cross,very thick,red] at (1,1.5*\y+1.125) {};
  \node[cross,very thick,red] at (-1,1.5*\y+1.125) {};
}

\node[below] at (1,-0.1) {$x$};
\node[below] at (-1,-0.1) {$-x$};

\node[right] at (0.1,1.125) {$\frac{i\beta}{2}$};
\node[right] at (0,-1.125) {$-\frac{i\beta}{2}$};
\node[right] at (0.1,3.375) {$\frac{3i\beta}{2}$};
\node[right] at (-.2,-3.5) {$-\frac{3i\beta}{2}$};
\draw[black] (-.1,1.125) -- (.1,1.125);
\draw[black] (-.1,-1.125) -- (.1,-1.125);
\draw[black] (-.1,3.375) -- (.1,3.375);
\draw[black] (-.1,-3.375) -- (.1,-3.375);
\draw[black] (-1,-.1) -- (-1,.1);
\draw[black] (1,-.1) -- (1,.1);
\end{tikzpicture}
    \caption{Schematic view of the analytic structure of $C(t,x)$ with $x\neq 0$. Crucially, at $x=0$ the branch cuts close up and there is no periodicity in imaginary time. } 
    \label{fig:placeholder}
\end{figure}
\indent We now argue that as long as $x$ is nonzero, this function can be analytically continued beyond the strip of analyticity, to a function with periodicity $C(t+i\beta,x)=C(t,x)$. To do this we use the edge of the wedge theorem. This theorem assumes that we have two regions of the complex plane whose boundaries intersect on an interval. Then if a function is analytic on both regions and agrees on their intersection, then it extends to an analytic function on their union. \\
\indent In our case the first region is the strip of analyticity $-\beta/2<\text{Im  }t<\beta/2$, and the second region is the next strip, $\beta/2<\text{Im  }t<3\beta/2$. Their boundaries intersect on the line $\text{Im }t=\beta/2$, so we need to show that $C(t+i\beta/2,x)=C(t-i\beta/2,x)$ for $|t|<x$. The first step is to use KMS to write $C(t-i\beta/2,x)=C(-t+i\beta/2,x)$. Next we have 
\begin{align}
C\left(-t+\frac{i\beta}{2},x\right)&=\text{Tr}(e^{-\beta H}\mathcal{O}(-t,\vec{x})\mathcal{O}(0,0))\notag\\
&=\text{Tr}(e^{-\beta H}\mathcal{O}(0,0)\mathcal{O}(-t,\vec{x}))\notag\\
&=C\left(t+\frac{i\beta}{2},x\right),
\end{align}
where we used the fact that the two operators are spacelike separated for $t<|x|$. \\
\indent This shows that $C(t,x)$ extends analytically to the next strip as a periodic function, and we can keep going to fill the full complex plane as in Figure \ref{fig:placeholder}. Here a crucial assumption was that $x$ is nonzero. In particular, at fixed spatial momentum we need to integrate over all $x$, so the analytic continuation of the correlator (if it exists) will not necessarily be periodic in imaginary time.

\section{Quasinormal modes in SYK}\label{qnmsyk}

\begin{figure}
    \centering
        \includegraphics[width=.7\textwidth]{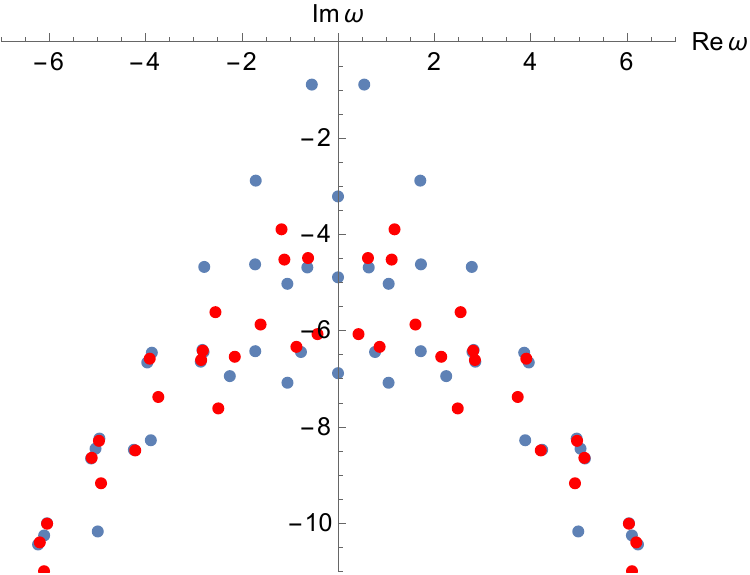}
        \caption{A pole-zero plot of the spectral function of $q=4$ SYK, with poles shown in blue and zeroes in red. Both poles and zeroes are contained within a tree at angle $\theta\approx \pi/3$ from the real axis. }
        \label{poleszeroes}
    \hfill
\end{figure}
The spectrum of QNMs in the SYK model was analyzed in \cite{Roberts:2018mnp,Dodelson:2024atp}. Here we present a computational method that leads to modest improvements in the result.\\
\indent The strategy taken in \cite{Dodelson:2024atp} was to fit the correlator to a finite sum of exponentials, 
\begin{align}\label{expfitting}
C(t)=\sum_{n=0}^{S-1}d_n e^{-i\omega_n t},
\end{align}
and then take the limit $S\to \infty$. The equation (\ref{expfitting}) can be solved on an evenly spaced grid $t_j=j\Delta t$, with $j=0,\ldots,2S-1$. This requires knowledge of the correlator for times $t<t_{\text{max}}\equiv (2S-1)\Delta t$. In the simplest approach, the correlator is computed by truncating the perturbative expansion at finite order, which implies that the maximum time sampled $t_{\text{max}}$ must be less than the radius of convergence $\beta_0/2$ of the perturbative expansion. \\
\indent In order to improve the result, we first extend the correlator to $t>\beta_0/2$ using Padé approximation. It is straightforward to check that the Padé representation of the correlator converges for all $t>0$. We then follow the procedure outlined in \cite{Dodelson:2024atp}, with $t_{\text{max}}$ chosen appropriately. In practice we found that the choice $t_{\text{max}}=2\beta_0$ gives an improvement over the results found in \cite{Dodelson:2024atp}. The resulting spectrum with $S=370$ is depicted in Figure \ref{poleszeroes}.

\bibliographystyle{JHEP}
\bibliography{mybib}
  
\end{document}